\documentclass[useAMS,usenatbib]{mnras}
\usepackage{graphicx}
\usepackage{multirow, bigstrut}%
\usepackage{epstopdf}%
\usepackage{color}%
\usepackage{soul}%
\usepackage{amssymb}
\usepackage{amsmath}
\usepackage[usenames,dvipsnames]{xcolor}%
\renewcommand{\vec}[1]{\mathbf{#1}}
 
\begin{document}

\title[2D shock/wind-cloud comparison]{A comparison of shock-cloud and wind-cloud interactions: The longer survival of clouds in winds}
\author[K. J. A. Goldsmith and J. M. Pittard] 
  {K. J. A. ~Goldsmith \thanks{pykjag@leeds.ac.uk} and J. M. ~Pittard\\
   School of Physics and Astronomy, University of Leeds, 
   Woodhouse Lane, Leeds LS2 9JT, UK}
   
\date{Accepted ... Received ...; in original form ...}

\pagerange{\pageref{firstpage}--\pageref{lastpage}} \pubyear{2016}
\newtheorem{theorem}{Theorem}[section]
\label{firstpage}

\maketitle

\begin{abstract}
The interaction of a hot, high-velocity wind with a cold, dense molecular cloud has often been assumed to resemble the evolution of a cloud embedded in a post-shock flow. However, no direct comparative study of these two processes currently exists in the literature. We present 2D adiabatic hydrodynamical simulations of the interaction of a Mach 10 shock with a cloud of density contrast $\chi=10$ and compare our results with those of a commensurate wind-cloud simulation. We then investigate the effect of varying the wind velocity, effectively altering the wind Mach number $M_{wind}$, on the cloud's evolution. We find that there are significant differences between the two processes: 1) the transmitted shock is much flatter in the shock-cloud interaction; 2) a low-pressure region in the wind-cloud case deflects the flow around the edge of the cloud in a different manner to the shock-cloud case; 3) there is far more axial compression of the cloud in the case of the shock. As $M_{wind}$ increases, the normalised rate of mixing is reduced. Clouds in winds with higher $M_{wind}$ also do not experience a transmitted shock through the cloud's rear and are more compressed axially. In contrast with shock-cloud simulations, the cloud mixing time normalised by the cloud-crushing time-scale $t_{cc}$ increases for increasing $M_{wind}$ until it plateaus (at $t_{mix} \simeq 25\, t_{cc}$) at high $M_{wind}$, thus demonstrating the expected Mach scaling. In addition, clouds in high Mach number winds are able to survive for long durations and are capable of being moved considerable distances.

\end{abstract}

\section{Introduction}
The flow of hot, high velocity gas through the interstellar medium (ISM) is known to play an important local role in star formation, and on much larger scales the formation and evolution of galaxies. The interaction of such flows with much cooler, dense clumps of gas (i.e. ``clouds'') can lead to the entrainment of cloud material. This shapes the morphology of the cloud and can ultimately cause the destruction of the cloud, altering the gas dynamics of the ISM (see \citet{Goldsmith16} for cases where the cloud is not destroyed on the usual dynamical time-scales). These interactions can inform our understanding of the nature of the ISM \citep[see e.g.][]{Elmegreen04, MacLow04, Scalo04, McKee07, Hennebelle12, Padoan14}, galaxy formation \citep[e.g.][]{Sales10}, and the evolution of supernova remnants (SNRs) and other diffuse sources e.g. \citep[e.g.][]{McKee77, Cowie81, White91, Dyson02, Pittard03}.

Observational studies have provided evidence of the interaction of hot flows with molecular clouds \citep[e.g.][]{Koo01, Westmoquette10}. High velocity winds and shocks in regions of star formation are capable of strongly affecting molecular clouds. For example, the B59 filament in the Pipe nebula is thought to be undergoing distortion by a wind \citep{Peretto12} and molecular cloud complexes in the Cygnus X region are being shaped by winds and radiation \citep{Schneider06}, whilst winds lead to the disruption, fragmentation, or dispersion of clouds such as the Rosette molecular cloud \citep{Bruhweiler10} (see also \citet{Rogers13} and \citet{Wareing17} for relevant numerical studies). Another effect of the interaction of a flow with a dense cloud is the entrainment of the cloud into the flow and acceleration of cloud material towards the flow's velocity. Several studies have revealed large outflow velocities from rapidly star-forming galaxies \citep[e.g.][]{Heckman00, Pettini01, Rupke02, Martin05, Martin12} and clouds have been typically observed at distances of a few kpc from the driving region \citep[e.g.][]{Soto12}. However, it has proved less easy to reconcile observations that clouds can travel distances on the order of 100 kpc without being destroyed \citep[e.g.][]{Turner14} by flows of such high velocity, and \citet{Scannapieco15} determined that in order to achieve these velocities clouds would need to be the size of entire galaxies. In addition to observations, shock-cloud interactions, in particular, have also been studied experimentally. For instance, the evolution of a sphere of dense material interacting with a laser-induced shock has been probed by X-ray radiography \citep{Klein03, Hansen07}.

The idealised case of a shock striking a spherical cloud was initially investigated numerically in the 1970s. \citet{Klein94} provided the first detailed 2D study of such interactions and examined the effects of varying the shock Mach number, $M$, and cloud density contrast, $\chi$, on the evolution of the cloud. Since then, numerous studies have been conducted in both 2D and 3D, many of which have included additional processes such as radiative cooling \citep[e.g.][]{Mellema02, Fragile04, Yirak10}, thermal conduction \citep[e.g.][]{Orlando05, Orlando08}, magnetic fields \citep[e.g.][]{MacLow94, Shin08, Johansson13, Li13}, turbulence \citep[e.g.][]{Pittard09, Pittard10, Pittard16b, Goodson17}, and multiple clouds \citep[e.g.][]{Poludnenko02, Aluzas12, Aluzas14}. Other numerical studies have considered how the nature of the interaction changes when the cloud is non-spherical \citep[e.g.][]{Xu95, Pittard16a, Goldsmith16}.

In addition to the large body of literature concerning shock-cloud interactions, many computational studies over the last two decades have considered the particular case of a hot, tenuous wind interacting with a cool, dense cloud (e.g. \citet{Klein94} briefly addressed the simple case of the 2D adiabatic interaction of a spherical cloud with a wind where the the initial shock has been removed - i.e. a cloud embedded within a post-shock flow). These studies have tended to focus on scenarios involving radiative cooling \citep[see e.g.][]{Marcolini05, Pittard05, Raga05, Raga07, Cooper08, Cooper09, Scannapieco15} or magnetic fields \citep[e.g.][]{Gregori99, Gregori00, McCourt15, Banda16}. 

The coupling of stellar feedback processes (winds, shocks from SNRs, etc.) with clouds can produce superficially-similar dynamical effects. \citet{Pittard09} noted that clouds with a high density contrast were able to survive the passage of a shock and would then be immersed in a post-shock flow that would resemble a wind with the same Mach number. Since the simulation of a hot, high-velocity wind can therefore be thought of as resembling a post-shock flow, many wind-cloud papers are highly pertinent to the shock-cloud scenario, and vice-versa. Although both wind-cloud and shock-cloud interactions have been well studied, there exists, to our knowledge, no direct comparison of the two processes in the literature. This, therefore, forms the motivation for our current work. In this paper we investigate a 2D hydrodynamical, adiabatic wind-cloud interaction and compare our results to those of a shock-cloud simulation using similar initial parameters. We then incrementally increase the velocity of the wind to increase its effective Mach number and explore the impact this has on the evolution of the cloud. A future paper will extend the analysis to include clouds with increased density contrasts. 

The outline of this paper is as follows: in Section 2 we introduce our numerical method and initial conditions. In Section 3 we present the results of our simulations. A brief discussion of the relevance of our work in terms of Mach scaling and the longevity of the cloud can be found in Section 4. Section 5 summarises and concludes.

\section{The numerical setup}
The Eulerian equations of inviscid flow are solved numerically for the conservation of mass, momentum, and energy:
\begin{equation}
\frac{\partial \rho}{\partial t} + \nabla \cdot (\rho \vec{u}) = 0,
\end{equation}

\begin{equation}
\frac{\partial \rho \vec{u}}{\partial t} + \nabla \cdot (\rho \vec{u} \vec{u}) + \nabla P =0 ,
\end{equation}

\begin{equation}
\frac{\partial E}{\partial t} + \nabla \cdot [(E+P) \vec{u}] = 0,
\end{equation}
respectively, where $\rho$ is the mass density, $\vec{u}$ is the velocity, $P$ is the thermal pressure, $\gamma$ is the ratio of specific heat capacities, and

\begin{equation}
E=\frac{P}{\gamma -1} + \frac{1}{2} \rho \vec{u}^{2}
\end{equation}
is the total energy density. In this study we limit ourselves to a purely hydrodynamical scenario, ignoring the effects of thermal conduction, radiative cooling, magnetic fields, background turbulence, and self-gravity. All computations were computed for an adiabatic, ideal gas, with $\gamma = 5/3$. The calculations in this study were performed using the \textsc{mg} hydrodynamical code which uses adaptive mesh refinement. The code solves a Riemann problem at each cell interface in order to determine the conserved fluxes for the time update, using piecewise linear cell interpolation. A linear solver is used in most instances, with the code switching to an exact solver where there is a large difference between the two states \citep{Falle91}. The scheme is second-order accurate in space and time.

A hierarchy of $n$ grid levels, $G^{0} \cdots G^{n-1}$, is used and two grids ($G^{0}$ and $G^{1}$) cover the entire computational domain, with finer grids being added where needed and removed where they are not. The amount of refinement is increased at points in the mesh where shocks or discontinuities exist, i.e. where the variables associated with the fluid show steep gradients. At these points, the number of computational grid cells produced by the previous level is increased by a factor of 2 in each spatial direction. Thus, fine grids are only utilised in regions where the flow is highly variable, with much coarser grids used where the flow is relatively uniform. Refinement and derefinement are performed on a cell-by-cell basis and are controlled by the differences in the solutions on the coarser grids at any point in space. Refinement occurs when there is a difference of more than 1 per cent between a conserved variable in the finest grid and its projection/prolongation from a grid one level down. If the difference in the two preceding levels falls to below 1 per cent, the cell is derefined. The time step on grid $G^{n}$ is $\Delta t_{0}/2n$, where $\Delta t_{0}$ is the time step on grid $G^{0}$. The effective resolution is taken to be the resolution of the finest grid and is given as $R_{cr}$, where `cr' is the number of cells per cloud radius in the finest grid. Each of our simulations was performed at an effective resolution of $R_{128}$. All length scales are measured in units of the cloud radius, $r_{c}$, where $r_{c}=1$, velocities are measured in units of the shock velocity through the ambient medium, $v_{b}$, the unit of density is taken to be the density of the ambient medium, $\rho_{amb}$, and the unit of pressure is the ambient pressure, $P_{amb}$. We impose no inherent scale on our simulations. Thus, our calculations can easily be applied to any physical scale required.

\subsection{Initial conditions}
To simulate a shock-cloud interaction, we consider a Mach 10 shock, initially located at $z=1$, interacting with a cloud of density contrast $\chi =10$ initially centred on the grid origin $r,\,z=(0,0)$ on a two-dimensional RZ cylindrically-symmetric grid. We retain these parameters for the simulations of a wind-cloud interaction but fill the entire domain external to the cloud with the post-shock flow, which mimics a mildly supersonic wind. We explore the effect of increasing the velocity of the flow, $v_{ps/wind}$ - effectively increasing the Mach number of the wind - on the evolution of the cloud. The numerical domain is set to be large enough so that the cloud is sufficiently mixed into either the post-shock flow or wind before reaching the edge of the grid. Table~\ref{Table1} details the grid extent for each of the simulations.

\begin{table}
\centering
\caption{The grid extent for each of the simulations (see \S 3 for the model naming convention). $M_{ps/wind}$ refers to the effective Mach number of the post-shock flow/wind. The unit of length is the initial cloud radius, $r_{c}$.}
\label{Table1}
\begin{tabular}{@{}lccc}
  \hline
 Simulation & $M_{ps/wind}$ &  $R$ & $Z$   \\
    \hline
c1shock & 1.36 &  $0 < R < 10 $& $-200 < Z < 5 $ \\
c1wind1 & 1.36 &  $0 < R< 20$ & $-400 < Z < 10$ \\
c1wind1a & 4.30 &  $0 < R < 20$ & $-400 < Z < 10$ \\
c1wind1b & 13.6 &  $0 < R < 20$ & $-500 < Z < 10$ \\
c1wind1c & 43.0 & $0 < R < 20$ & $-500 < Z < 10$ \\
  \hline
 \end{tabular}
\end{table}
 
The simulated cloud is assumed to have sharp edges, which maximises the growth of KH instabilities and sets a lower limit to the cloud's lifetime \citep[see e.g.][]{Nakamura06, Pittard16b}. The shock-cloud simulation is described by the sonic Mach number of the shock, $M_{shock}$, and the density contrast between the cloud and the stationary ambient medium, $\chi$. The cloud is initially in pressure equilibrium with its surroundings. The Mach number of the post-shock flow/wind is defined as
\begin{equation}
M_{ps/wind} = \frac{v_{ps/wind}}{c_{ps/wind}} \, ,
\end{equation}
where $c_{ps/wind}$, the adiabatic sound speed of the post-shock flow/wind, is given by $c_{ps/wind} = \sqrt{\gamma \frac{P_{ps/wind}}{\rho_{ps/wind}}}$.

For the $M=10$ shock-cloud simulation, the post-shock density, pressure, and velocity are $\rho_{ps}/\rho_{amb} = 3.9$, $P_{ps}/P_{amb} = 124.8$, and $v_{ps}/v_{b} = 0.74$, respectively. In model $c1wind1$, the cloud is completely surrounded by the post-shock flow conditions used in model $c1shock$. It thus interacts with a flow which has the same density, pressure, and velocity as the post-shock material in model $c1shock$. The cloud is thus under-pressured compared to the surrounding flow, but at exactly the same pressure as in the shock-cloud simulation.\footnote{This is a slightly different set-up, therefore, compared to most previous wind-cloud investigations, but is necessary for a more direct comparison to shock-cloud interactions.} The Mach number of this flow/wind (with respect to the cloud) is $M_{ps/wind}=1.36$. In the remaining wind models, the velocity of the wind is increased by factors of $\sqrt{10}$, $\sqrt{100}$, and $\sqrt{1000}$ in models $c1wind1a$, $c1wind1b$, and $c1wind1c$, respectively. This results in an increase in the Mach number of the wind. Values for $M_{wind}$ for each of these simulations are given in Table~\ref{Table1}. However, the sound speed of the wind remains the same throughout.

\subsection{Global quantities}
Various diagnostic quantities are used to follow the evolution of the interaction \citep[see][]{Klein94, Nakamura06, Pittard09, Pittard16b}, including the ablation and mixing of the cloud, as well as the acceleration of the cloud by the flow. These quantities include the cloud mass ($m$), mean velocity in the $z$ direction ($\langle v_{z} \rangle$), and velocity dispersions along each orthogonal axis (e.g. $\delta v_{z}$). Averaged quantities $\langle f \rangle$ are constructed by
\begin{equation}
\langle f \rangle = \frac{1}{m_{\beta}} \int_{\kappa \geq \beta} \kappa \rho f \; \mathrm{d}V ,
\end{equation}
where $m_{\beta}$, the mass which is identified as being part of the cloud, is given by
\begin{equation}
m_{\beta} = \int_{\kappa \geq \beta} \kappa \rho \; \mathrm{d}V .
\end{equation}
An advected scalar, $\kappa$, is used to distinguish between the cloud and ambient material in the flow, allowing the whole cloud to be tracked. $\kappa$ has an initial value of 1.0 within the cloud, and is zero for the ambient material. $\beta$ is the threshold value, and integrations are performed over cells where $\kappa \geq \beta$. Two related sets of quantities can thus be investigated: setting $\beta=0.5$ explores the densest regions of the cloud and its associated fragments (hereafter subscripted as ``core"). Setting $\beta=2/\chi$ explores the entire cloud, including regions where cloud material is well mixed with the ambient flow (hereafter subscripted as ``cloud"). We define motion in the direction of wind/shock propagation as ``axial'' (the wind/shock propagates in the negative $z$ direction), whilst motion perpendicular to this is termed ``radial''.

In order to measure the shape of the cloud, the effective radii of the cloud in the radial ($a$) and axial ($c$) directions are defined as
\begin{equation}
a = \left(\frac{5}{2} \langle r^{2}\rangle \right)^{1/2}, \; \; \; c = [5(\langle z^{2}\rangle - \langle z\rangle^{2})]^{1/2} \, .
\end{equation}

\subsection{Time-scales}
For the shock-cloud simulation, we use the characteristic time-scale for a cloud to be crushed (the ``cloud-crushing time") given by \citet{Klein94}:
\begin{equation}
t_{cc} = \frac{\sqrt{\chi} \,r_{c}}{v_{b}}\, .
\end{equation}
For the wind-cloud simulations we redefine this time-scale in terms of the velocity of the wind past the cloud ($v_{ps/wind}$):
\begin{equation}
t_{cc} = \frac{0.74\, \sqrt{\chi}\,r_{c}}{v_{ps/wind}}\, ,
\label{eqn11}
\end{equation}
where $v_{ps/wind}=0.74\,v_{b}$ (the constant 0.74 is specific to the Mach 10 shock simulation against which the wind simulations are compared).\footnote{Note that in some wind-cloud studies, $t_{cc}$ is defined slightly differently \citep[e.g.][]{Jones96, Banda16}.} Since this time-scale is dependent on the cloud density contrast and the speed of the flow, those simulations that share the same value of $\chi$ and $v_{ps/wind}$ (e.g. $c1shock$ and $c1wind1$) have identical values of $t_{cc}$. However, as the wind Mach number is increased, the value of $t_{cc}$ decreases because of its dependence on $v_{ps/wind}$. Values for the cloud crushing time for each simulation are given in Table~\ref{Table3}. Several other time-scales are also available. For example, the ``drag time'', $t_{drag}$, is the time taken for the average cloud velocity relative to the post-shock flow or wind to decrease by a factor of $e$ (i.e. the time when the average cloud velocity $\langle v \rangle_{cloud}\,=(1-1/e)\,v_{ps/wind}$); the ``mixing time'', $t_{mix}$, is the time when the cloud core mass is half that of its initial value, and the cloud ``lifetime'', $t_{life}$, is the time taken for the cloud core mass to reach one per cent of its initial value.

Time zero in our calculations is taken to be the time when the shock is level with the leading edge of the cloud, in the shock-cloud case, whilst for the wind-cloud case the simulation begins with the cloud immediately surrounded by the flow.
    
\section{Results}
In this section we present the results from our various simulations. We begin with a brief examination of the interaction of a shock with a cloud in terms of its morphology and then, maintaining the same initial parameters, compare this to the interaction of a wind with a cloud. We then consider in detail the interaction of clouds with winds of increasing Mach number.  

At the end of this section we consider the impact of the interaction on various global quantities. We adopt a naming convention for each simulation such that $c1shock$ refers to a shock-cloud simulation with $\chi = 10$. Models with $wind1a-c$ in their title indicate wind-cloud interactions of increasing wind Mach number.

\subsection{Stages}
The purely adiabatic evolution of a cloud struck by a shock propagating in the $-z$ direction is characterised by four main stages \citep[see e.g.][]{Pittard16b}: (1) the cloud is struck by the shock, causing a transmitted shock to travel at a velocity $v_{s}$ through the cloud, while a bow shock (or bow wave) is formed upstream and the incident shock diffracts around the cloud; (2) the cloud undergoes compression in the $z$ direction (on the whole) by both the transmitted shock and also a shock driven into the back of the cloud due to a dramatic pressure jump as the external shock is focussed onto the axis; (3) the cloud reaches the expansion stage where, under high pressure, it expands in the radial and axial directions; and (4) the cloud is finally destroyed and mixed with the post-shock flow.

In the case of a wind-swept cloud, stages 1-4 remain essentially the same. However, since the cloud immediately begins interacting with the flow, \citet{Banda16} divided the stages for a wind-cloud scenario thus: 1) compression, including the transmission and reflection of shocks within, and external to, the cloud; 
2) stripping; 3) expansion; and 4) break-up. They noted that the stripping phase (when cloud material begins to flow downstream and wraps around the cloud, converging on the axis behind the cloud) occurs at all times, but is more dynamically important up to $t \approx 1.3 \, t_{cc}$.

\subsection{Shock-cloud interaction}
We begin by examining the morphology of the interaction for the shock-cloud scenario, where $M=10$ and $\chi=10$ (simulation $c1shock$). The shock is initially located at $z=1$ (i.e. level with the leading edge of the cloud).

\begin{figure*}
\centering
\begin{tabular}{c}
\includegraphics[width=150mm]{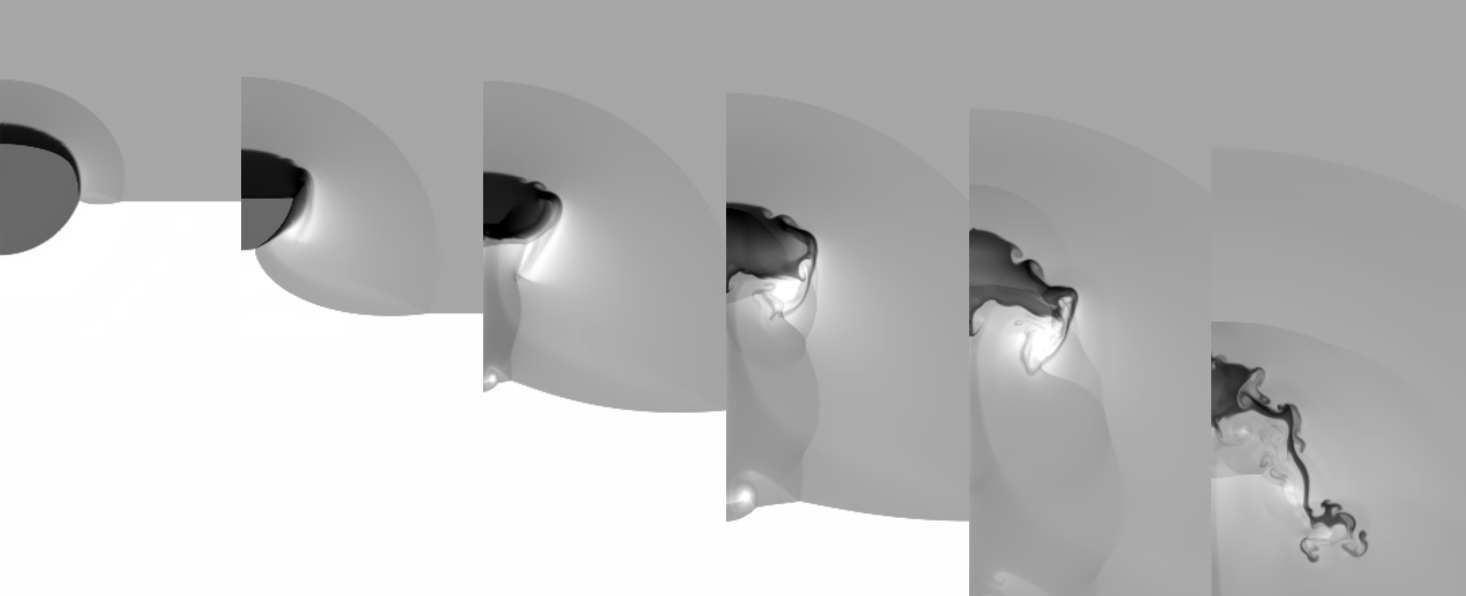}\\
\includegraphics[width=150mm]{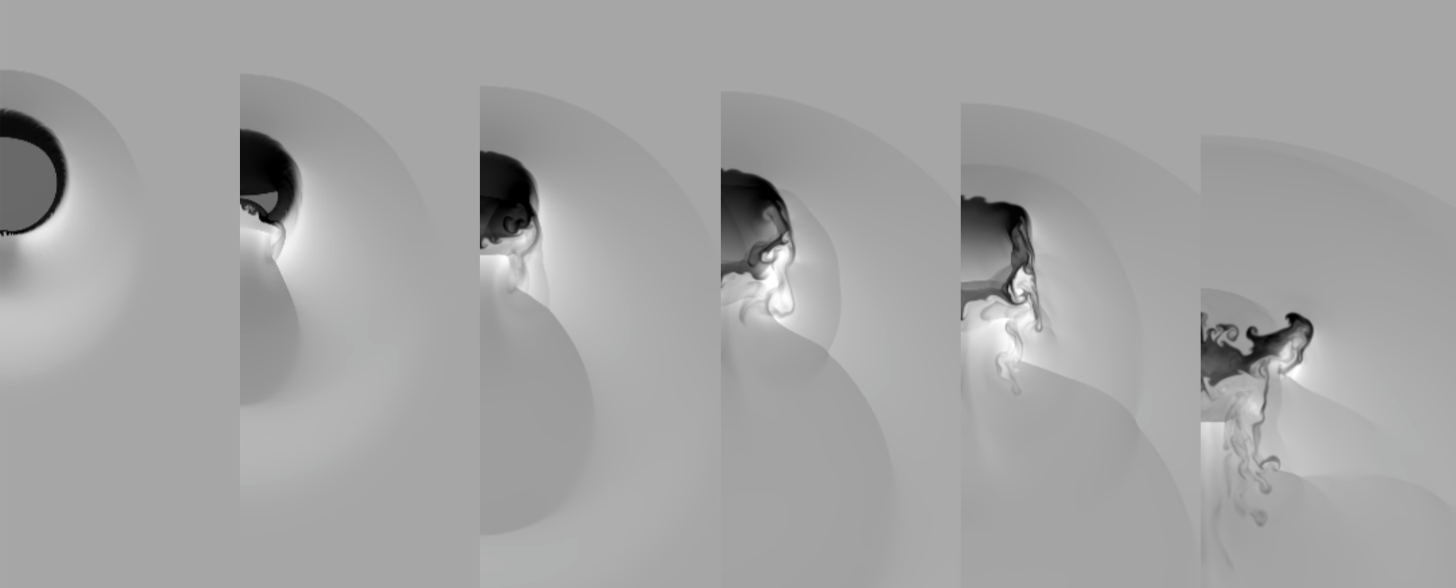}\\
\end{tabular}
\caption{The time evolution of the logarithmic density for models (top) $c1shock$ and (bottom) $c1wind1$. The greyscale shows the logarithm of the mass density, from white (lowest density) to black (highest density). The density in this and subsequent figures has been scaled with respect to the ambient density, so that a value of 0 represents the value of $\rho_{amb}$ and 1 represents $10 \times \rho_{amb}$, and the density scale used for this figure extends from 0 to 1.7. The evolution proceeds left to right with $t=0.43 \,t_{cc}$, $t=0.82 \,t_{cc}$, $t=1.2 \,t_{cc}$, $t=1.6 \,t_{cc}$, $t=2.0\,t_{cc}$, and $t=3.3 \,t_{cc}$. The $r$ axis (plotted horizontally) extends $3\,r_{c}$ off-axis in each plot. All frames in the top and bottom sets show the same region ($-5 < z < 2$, in units of $r_{c}$) so that the motion of the cloud is clear. Note that in this and similar figures the $z$ axis is plotted vertically, with positive towards the top and negative towards the bottom.}
\label{Fig1}
\end{figure*}

Figure~\ref{Fig1} (top panels) provides logarithmic density plots of the $rz$ plane as a function of time for the shock-cloud case. The evolution of the cloud broadly follows the above stages. The shock initially strikes the cloud on its leading edge, sending a transmitted shock through the cloud whilst the external shock is bent around the edge of the cloud as it moves downstream. The external shock becomes level with the centre of the cloud at $t\simeq 0.32 \,t_{cc}$. A bow shock is visible upstream of the cloud. The first three upper panels of Fig.~\ref{Fig1} relate approximately to the first two stages of evolution, which lasts until $t \simeq t_{cc}$. The external shock sweeps around the cloud and becomes focussed on the $r=0$ axis. A region of higher pressure forms downstream behind the cloud due to the convergence of this shock on the axis and this serves to drive secondary shocks back through the cloud towards its leading edge. These secondary shocks create additional waves and shocks upstream of the cloud (note the faint secondary shock front just ahead of the cloud in the upper panel at $t=2.0 \, t_{cc}$ in Fig.~\ref{Fig1}) when they exit the leading edge of the cloud, accelerating as they do so.

At $t\simeq 1.6\,t_{cc}$ the transmitted shock has exited the back of the cloud and accelerates into the downstream gas. This action initiates a rarefaction wave which propagates in the upstream direction. The secondary shocks deposit vorticity as they progress back through the cloud. This deposition begins to disrupt the smooth morphology of the cloud, forcing the right-hand edge of the cloud upwards and leading to a modest expansion of the cloud in the transverse direction. At the same time, a supersonic vortex ring forms downstream of the cloud on the $r=0$ axis. In a similar manner to e.g. \citet{Pittard09} and \citet{Pittard16b}, the cloud exhibits a low-density interior surrounded by a thick, high-density shell (see upper panel at $t=1.6 \, t_{cc}$ in Fig.~\ref{Fig1}). At $t\simeq 2.0\,t_{cc}$, the shell begins to collapse. Cloud material is now ablated by the surrounding flow and shear instabilities at the side of the cloud result in a ``rolling-up'' of cloud material in the transverse direction - over time this becomes shredded into long strands by the action of KH instabilities on the surface of the cloud. In addition, there is some circulation of the flow on the axis behind the cloud which serves to strip material from the rear of the cloud, allowing it to mix in with the flow. After $t\simeq 3.3\,t_{cc}$, a long, turbulent wake forms on the axis downstream of the cloud, and the cloud is quickly ablated.

\subsection{Wind-cloud interaction}
\subsubsection{Comparison of wind-cloud and shock-cloud interactions}
Figure~\ref{Fig1} (bottom panels) shows logarithmic density plots of the $rz$ plane as a function of time for the wind-cloud case with $M_{wind}=1.36$ and $\chi=10$ (simulation $c1wind1$). The velocity, density, and pressure of the wind are exactly the same as the post-shock values in simulation $c1shock$ (i.e. the cloud is surrounded by `post-shock' material). Hence, the density jump between the cloud and the wind is given by $\chi/3.9$ (see \S 2.1).

The morphology of the cloud and its evolution shares some broad similarities with the shock-cloud case (e.g. both clouds form dense shells surrounding lower density interiors, both are squeezed in the radial direction, and both are eventually drawn into long, filamentary wakes in the axial direction), but there are also some key differences.

Firstly, there are clear differences in the behaviour of the external medium. Since the simulation begins with the marginally supersonic wind completely surrounding the cloud, a small lower-density, lower-pressure region is immediately formed on the axis downstream of the cloud \citep[as also noted by][]{Marcolini05, Banda16}. This feature is not present in the shock-cloud case and is formed by the initial motion of the wind removing gas from around the rear of the cloud. The low-pressure region is eventually carried downstream of the cloud, allowing an area of higher pressure to form behind the cloud (though not in quite the same manner as in the $c1shock$ simulation).

Secondly, whilst the cloud is strongly compressed into the shape of an oblate spheroid in the shock-cloud case, the cloud in the wind-cloud case suffers much less compression in the axial direction, particularly during the initial stages of the interaction, and maintains a more rounded shape. While the leading edge of the cloud undergoes much less compression compared to the shock case, the rear of the cloud is clearly being pushed upwards by the action of a shock driven into the back of the cloud. Plots of the logarithmic pressure (not shown) indicate that a region of high pressure occurs at the leading edge of the cloud in both models, while the back of the cloud remains at a relatively lower pressure in model $c1wind1$ compared to $c1shock$. In their study of a wind-cloud interaction with $M_{wind}=10$ (i.e. a higher wind Mach number than used in our model $c1wind1$), \citet{Schiano95} noted generally that when a free-flowing wind encounters a 2D spherical cloud and passes through the bow shock, the wind is compressed, decelerated, heated, and channelled around the cloud. As the shocked gas is accelerated around the periphery of the cloud and rejoins the wind flow along the cloud flanks, the gas pressure is lowered, and there is therefore a commensurate decrease in cloud pressure with increasing distance from the cloud apex; this is similar to the situation in model $c1wind1$.

There are also clear differences between the two simulations in terms of the initial transmitted shock driven through the cloud. In model $c1shock$, the shock is reasonably flat as it progresses through the cloud, whereas it is much less flat in model $c1wind1$ (cf. both panels at $t=0.43 \, t_{cc}$ and $t=0.82 \, t_{cc}$ in Fig.~\ref{Fig1}) and curves around the edge of the cloud. As in the shock-cloud case, secondary shocks driven back into the cloud lead to the formation of shocks/waves upstream of the cloud, though in model $c1wind1$ these are slightly more pronounced (e.g. at $t=2.0\,t_{cc}$).

At $t= 2.8\,t_{cc}$, the cloud, which has developed a dense shell surrounding a less dense interior, collapses at a slightly later time than in the shock-cloud case. Eventually the cloud takes on a very similar morphology to that in model $c1shock$ from $t\simeq 3.3\,t_{cc}$ onwards, when it is drawn into a long wake in the axial direction (not shown).

\subsubsection{Effect of increasing $M_{wind}$ on the evolution}

\begin{figure*}
\centering
\begin{tabular}{c}
\includegraphics[width=150mm]{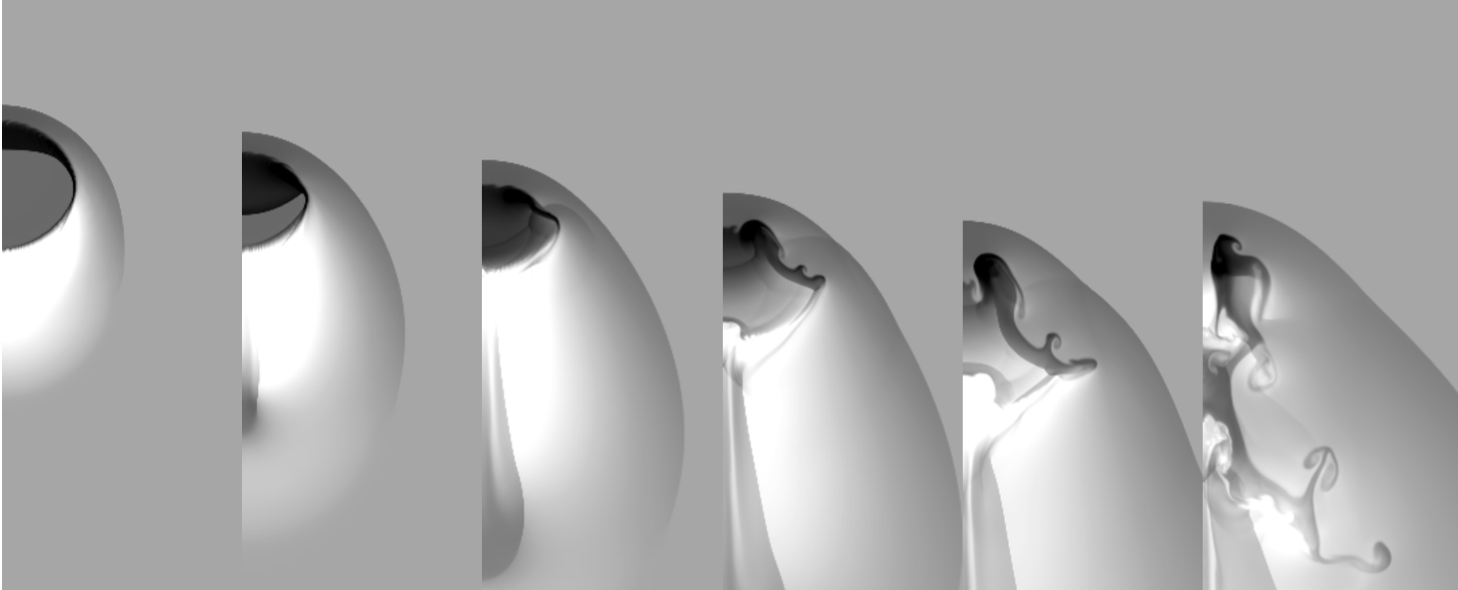} \\
\includegraphics[width=150mm]{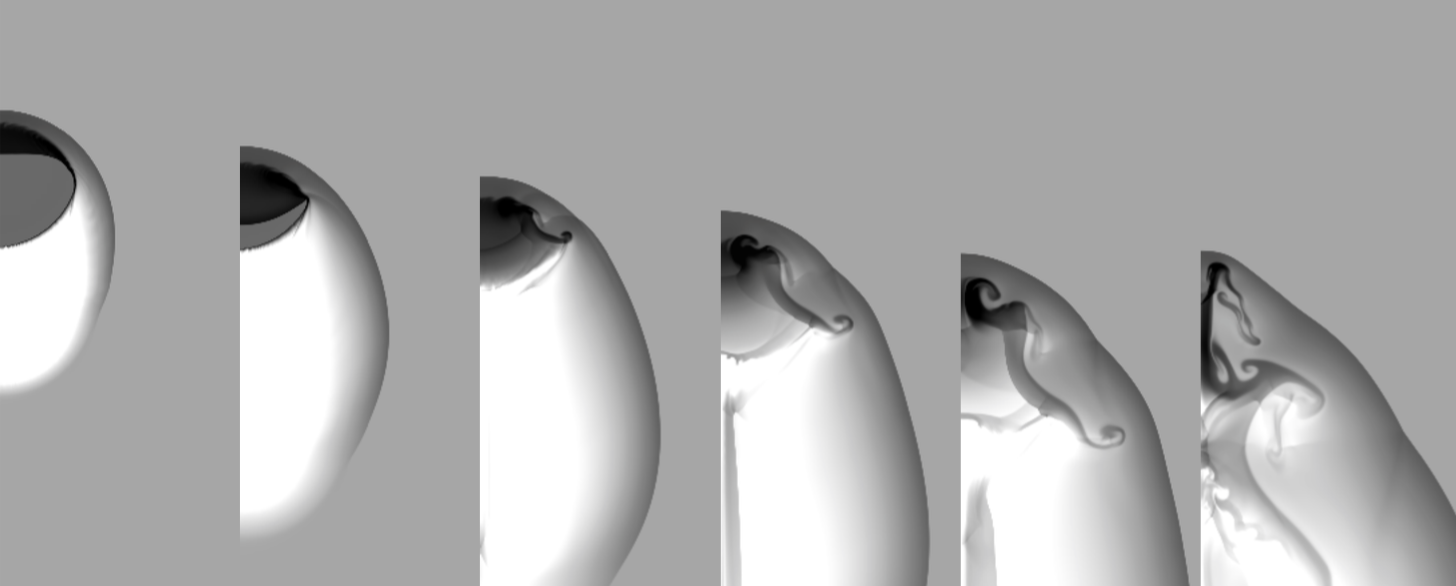} \\
\includegraphics[width=150mm]{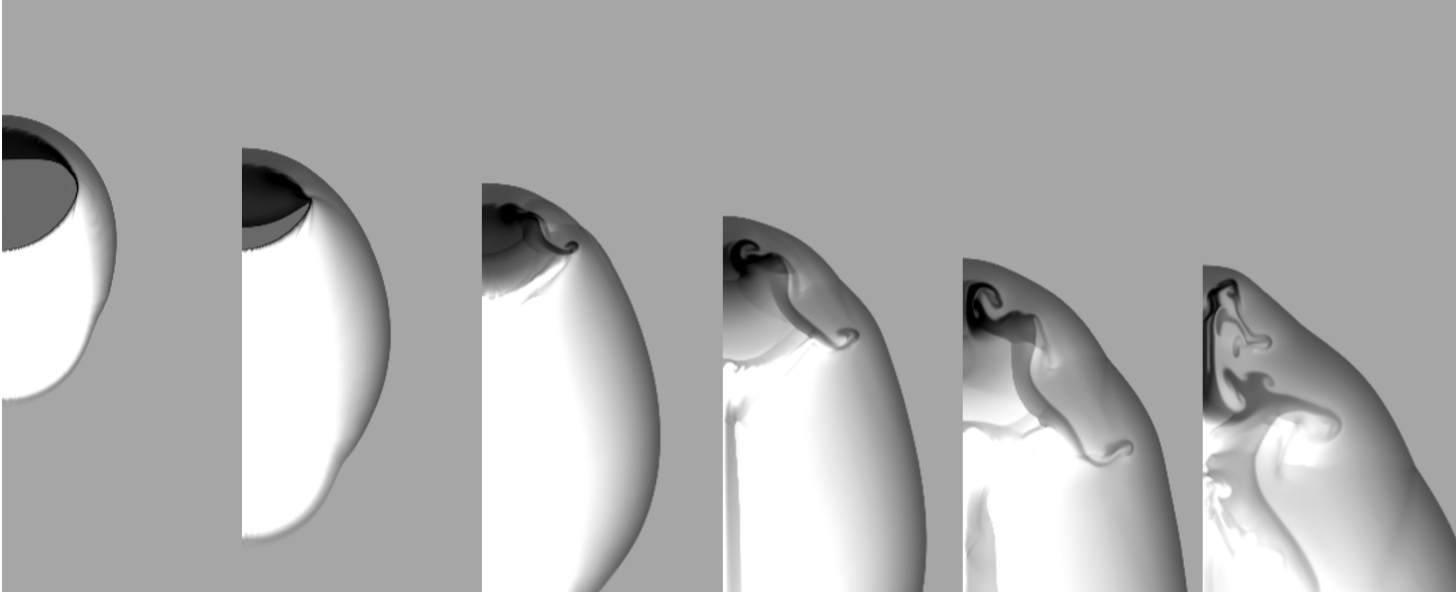} \\
\end{tabular}
\caption{The time evolution of the logarithmic density for models (top) $c1wind1a$, (middle) $c1wind1b$, and (bottom) $c1wind1c$. The greyscale shows the logarithm of the mass density, scaled with respect to the ambient medium. The density scale used in this figure extends from 0 to 1.7. The evolution proceeds left to right with $t=0.7 \,t_{cc}$, $t=1.3 \,t_{cc}$, $t=1.9 \,t_{cc}$, $t=2.6 \,t_{cc}$, $t=3.2\,t_{cc}$, and $t=5.2 \,t_{cc}$. The $r$ axis (plotted horizontally) extends $3\,r_{c}$ off-axis in each plot. The first 5 frames in each set show the same region ($-5 < z < 2$, in units of $r_{c}$) so that the motion of the cloud is clear. The displayed region is shifted in the last frame in each set ($-7 < z < 0$) in order to more fully show the cloud.}
\label{Fig2}
\end{figure*}

Figure~\ref{Fig2} shows the time evolution of the logarithmic density for models $c1wind1a$, $c1wind1b$, and $c1wind1c$, where the wind has an increasing Mach number ($M_{wind} =$ 4.3, 13.6, and 43, respectively). As can be seen from a comparison between Fig.~\ref{Fig2} and the lower panels of Fig.~\ref{Fig1}, there are a large number of differences between these simulations and $c1wind1$ (where $M_{wind}=1.36$). 

Firstly, as the effective Mach number of the wind increases, the region of low pressure behind the cloud becomes a very low-pressure cavity, is highly supersonic, and expands rapidly in the direction of wind propagation, becoming elongated as it does so. Unlike the initial wind-cloud interaction described above ($c1wind1$), these cavities do not move away from the rear of the cloud, and because they are of a much lower pressure than the region in $c1wind1$ they are far more pronounced.

Secondly, a transmitted shock moves inwards from the back of the cloud in $c1wind1$ but not in the higher $M_{wind}$ simulations. Whilst the wind flow around the cloud in model $c1wind1$ is focussed around the cloud flank and onto the $r=0$ axis, in models $c1wind1b$ and $c1wind1c$ the flow is much more linear and suffers very little deflection at the back of the cloud. Because of this, there is no dramatic pressure jump behind the cloud and this helps prevent a transmitted shock being driven into the back of the cloud.

Thirdly, it is noticeable that the density jump at the bow shock and the stand-off distance between the bow shock and the leading edge of the cloud both change according to the Mach number (see Table~\ref{Table2}). As the Mach number of the wind increases, the density jump increases towards the high Mach number limit and the stand-off distance between the bow shock and cloud decreases (see \citet{Farris94} for a discussion of the factors affecting the stand-off distance). The higher post-shock pressure behind the bow shock causes the leading edge of the cloud to be pushed slightly further downstream in the higher $M_{wind}$ simulations, compared to $c1wind1$. The normalised velocity of the shocked gas around the edge of the cloud is also reduced due to the higher compression at the bow shock. The nature of the transmitted shock propagating through the cloud also changes, becoming initially much flatter as the Mach number increases, more akin to the shock-cloud case. All of this serves to compress the cloud in the axial direction, lending it an oblate spheroid shape similar to the cloud in model $c1shock$, rather than the more rounded morphology evident in model $c1wind1$. Although the shape of the cloud in all the wind-cloud simulations with higher values of $M_{wind}$ is similar, compared to that in model $c1wind1$, it is noticeable that the cloud in model $c1wind1a$ becomes more kinked on its leading edge with the kink resembling the beginnings of a finger of cloud material moving in the $+z$ direction, and that the development of this kink is different, compared to models $c1wind1b$ and $c1wind1c$ where the kink is more curled and resembles a KH instability (cf. final two panels in each set of Fig.~\ref{Fig2}). The effect of this kink on the lifetime of the cloud is discussed in \S3.4.1.

It should be noted that the cloud morphology and statistics in simulations $c1wind1b$ and $c1wind1c$ are very similar \citep[as expected from Mach scaling - cf.][]{Klein94, Pittard10}.

\begin{figure}
\centering
\begin{tabular}{c}
\includegraphics[width=70mm]{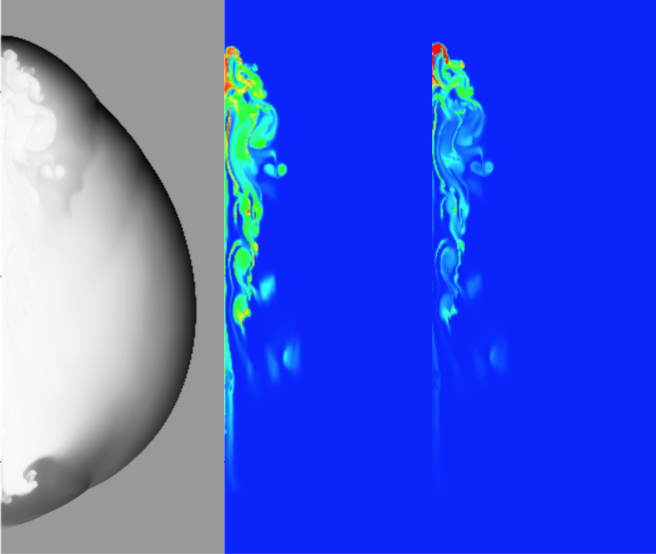} \\
\includegraphics[width=70mm]{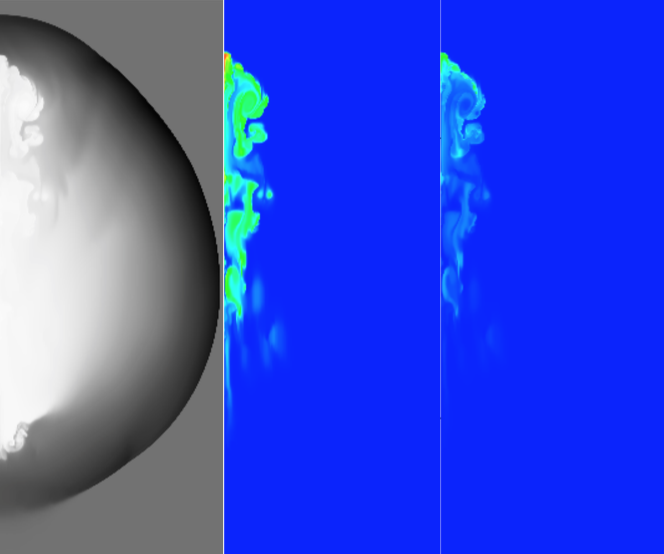} \\
\end{tabular}
\caption{The time evolution of the linear density (left), advected scalar $\kappa$ which identifies only the cloud material (middle), and advected scalar $\times$ linear density which allows the density of only the cloud to be shown (right) for model $c1wind1c$. The greyscale shows the mass density, scaled with respect to the ambient medium. The density scale used in the left-hand panels of this figure extends from 0 to 9.7 in the upper panels and 0 to 7.0 in the lower panels. The colour scale in the middle frames extends from dark blue (ambient material) to red (cloud material). The scale used in the right-hand panels extends from 0 to 1 (the ambient medium has a density of 1, but an advected scalar of 0, in this plot). All of the top panels are at $t=46.0 \,t_{cc}$, whilst all the bottom panels are at $t=101.4 \,t_{cc}$. The $r$ axis (plotted horizontally in each frame) extends $12\,r_{c}$ off-axis in the top set of frames and $16 \,r_{c}$ off-axis in the bottom set of frames. All frames in the top set show the same region ($-115 < z < -85$, in units of $r_{c}$) whilst all frames in the bottom set show $-250 < z < -210$.}
\label{Fig3}
\end{figure}

Figure~\ref{Fig3} shows the density, advected scalar $\kappa$, and advected scalar $\times$ density for model $c1wind1c$ at late times (i.e. $t=46 \,t_{cc}$ and $t=101\, t_{cc}$). It can clearly be seen that the cloud has yet to be smoothed out into the flow and shows some evidence of structure along with a distinct cloud edge. Compared with the lower panels in Fig.~\ref{Fig2}, which show the cloud during the initial stages of the evolution, the cloud in Fig.~\ref{Fig3} has expanded supersonically into the flow and formed a tail-like structure. Although the cloud is not highly dense at late times, we can infer that it, nonetheless, shows evidence of long-term survival, something that has not been observed in previous wind-cloud studies.

\begin{table}
\centering
\caption{Values of the density jump and bow shock stand-off distance (in units of $r_{c}$) for each of the simulations.}
\label{Table2}
\begin{tabular}{@{}lcc}
  \hline
 Simulation & Density jump &  Stand-off distance   \\
    \hline
c1shock & 1.53  &  1.72  \\
c1wind1 & 1.53 &  1.72   \\
c1wind1a & 3.44  &  1.32   \\
c1wind1b & 3.94  &  1.28   \\
c1wind1c & 3.99  & 1.28   \\
  \hline
 \end{tabular}
\end{table}

\subsection{Statistics}
We now explore the evolution of various global quantities of the interaction for both the shock-cloud and wind-cloud models. Figure~\ref{Fig4} shows the time evolution of these key quantities, whilst Table~\ref{Table3} lists various time-scales taken from these simulations. The following subsections present a more detailed discussion of these statistics.

\begin{table}
\centering
 \caption{A summary of the cloud-crushing time, $t_{cc}$ for a cloud with $\chi=10$ and $r_{c}=1$ (see Eqn.~\ref{eqn11} re the calculation of $t_{cc}$), and key time-scales, in units of $t_{cc}$, for the simulations investigated in this work. Note that the value for $t_{drag}$ given here is calculated using the definition given in \S 2.3, in comparison to the values shown in Fig.~\ref{Fig6} which were calculated using the definition given in \citet{Pittard10} in order to compare with the values of $t_{drag}$ presented in that paper.}
 \label{Table3}
 \begin{tabular}{@{}lcccc}
  \hline
  Simulation & $t_{cc}$ & $t_{drag}$ & $t_{mix}$ & $t_{life}$ \\
  \hline
c1shock  & 0.233 & 2.35	& 6.72 & 23.0 \\
c1wind1  & 0.233 & 3.34	& 6.12 & 12.9 \\
c1wind1a  & 0.074 & 3.88	& 13.3 & 35.7  \\
c1wind1b  & 0.023 & 3.78	& 23.5 & 96.9 \\
c1wind1c  & 0.0074 & 4.28 & 25.6 & 136.0 \\
  \hline
 \end{tabular}
\end{table}

\subsubsection{Cloud mass}
Panel (a) of Fig.~\ref{Fig4} shows the time evolution of the core mass, $m_{core}$. The core mass decreases as a result of cloud material being ablated by, and mixed into, the surrounding flow. It is clear that models $c1shock$ and $c1wind1$ share a similar trend in terms of their rate of mass loss, until around two fifths of their core mass has been lost (both models have a much steeper rate of mass loss, at least until $t\approx 8 \, t_{cc}$, than the models with higher values of $M_{wind}$). This is surprising considering that the clouds in these simulations initially evolve very differently; for example, the passage of the shock through the cloud, the degree of compression of the cloud, and the presence or otherwise of a low-pressure region behind the cloud are different between the two simulations, leading to a difference in cloud morphology. In contrast, models $c1wind1b$ and $c1wind1c$ display very shallow curves which are almost coincident. This reduced rate of mass loss may be due to the lack of a transmitted shock being driven into the back of the cloud (in contrast to models $c1shock$ and $c1wind1$), as well as reduced circulation of the flow on the axis behind the cloud as $M_{wind}$ increases. In addition, the normalised wind velocity (in units of $v_{wind}$) is reduced around the cloud flank due to the increased compression at the bow shock. Thus, there is less stripping of material from the rear of the cloud compared to lower $M_{wind}$ simulations. 

Interestingly, model $c1wind1a$ appears to bridge the two groups: it is initially slow to lose mass (as per the other high $M_{wind}$ models), but between $t \approx 10-20\, t_{cc}$ its rate of mass loss gradually becomes comparable to that of the $c1shock$ and $c1wind1$ simulations. In simulations $c1wind1a-c$, a prominent ``kink'' develops on the leading edge of the cloud; this feature is not evident in Figure 2 of \citet{Pittard16b} but the difference may be attributable to the fact that we used a hard edge to our cloud which is more conducive to the growth of such instabilities. A similar kink is present in the adiabatic cloud modelled in \citet{Marcolini05}. This kink allows a greater expansion of the cloud in the radial direction (i.e., $a_{cloud}$ increases) at later times compared to models $c1shock$ and $c1wind1$. The kink develops differently between models $c1wind1a$ and $c1wind1b/c$, and the radial expansion of the cloud in model $c1wind1a$ occurs earlier than that of the latter two models. This means that the subsequent mixing and ablation of cloud material by the flow takes place earlier than in models $c1wind1b$ and $c1wind1c$.

\citet{Pittard16b} showed that the mixing time, $t_{mix}$, for a spherical cloud struck by a Mach 10 shock was $\approx 6 \, t_{cc}$ and increased as the value of the shock Mach number was reduced. Table~\ref{Table3} shows that the two models with similar initial parameters ($c1shock$ and $c1wind1$) have roughly similar mixing times. However, for \textit{winds} of increasing Mach number the value of $t_{mix}$ \textit{increases} until near to the high Mach number limit (when $M_{ps/wind} \gtrsim 10$). As before, this is due to the less effective stripping of cloud material by the flow around the edge of the cloud as $M_{wind}$ increases. It is surprising, however, to find that the normalised mixing time is 5 times longer for clouds in winds than for clouds hit by shocks in the high Mach number limit.

\subsubsection{Cloud velocity}
Figure~\ref{Fig4}(b) shows the mean velocity of the cloud in the direction of propagation of the shock/wind, normalised by the post-shock/wind velocity. The clouds in models $c1shock$ and (from $t\approx 4\, t_{cc}$) $c1wind1$ show slightly faster acceleration towards the asymptotic velocity, with the cloud in $c1wind1$ being accelerated to the velocity of the background flow much more quickly than in the other wind simulations. In addition, in model $c1shock$ (and to a much lesser extent $c1wind1$), the cloud exhibits a ``two-stepped'' acceleration at $t \approx 4 \, t_{cc}$. This coincides with the beginning of a `plateau' region. At this point, the cloud undergoes significant stretching in the axial direction until $t \approx 8\, t_{cc}$ (the approximate end of the plateau region), when most of the core material has been ablated and the remaining less dense and filamentary structure is again accelerated by the flow up to the asymptotic velocity. 

The acceleration of the cloud in model $c1wind1a$ is initially smooth until $t \approx 15\, t_{cc}$ (at which point the cloud begins to form long strands), but then fluctuates slightly about the velocity of the wind. The clouds in models $c1wind1b$ and $c1wind1c$ undergo the smoothest acceleration because of the reduction in the growth of turbulent instabilities on the cloud surface, and again are almost identical in behaviour (due to Mach scaling).

\subsubsection{Centre of mass of the cloud}
The distance travelled by the cloud before it becomes fully mixed into the flow is reflected by the movement of the cloud centre of mass. The time evolution of the position of the centre of mass of the cloud in the $z$ direction, normalised by the initial radius of the cloud, is given in Fig.~\ref{Fig4}(c). It is clear that the post-shock flow or wind can transport cloud material over large distances. Up until $t\approx2\, t_{cc}$, there is not a great deal of movement in the direction of the flow (the centre of mass has only moved $0.8-1.8 \, r_{c}$). However, by $t=12\, t_{cc}$ the clouds have been displaced by $15-20$ times the initial cloud radius. Over much longer time spans (e.g. up to $t = 30\, t_{cc}$, as in Fig.~\ref{Fig4}(c)), the cloud displacement shows greater variation between models, with the centre of mass of the cloud in models $c1shock$ and $c1wind1$ showing considerably more movement. However, there is much less variety in displacement among all the higher wind Mach number simulations, indicating that movement in the axial direction is not strongly dependent upon $M_{wind}$ in these cases (as expected with Mach scaling). Figure~\ref{Fig4}(c) also shows the displacement of each cloud at $t=t_{mix}$. Clearly, the distance over which the cloud has moved by the time its core mass has been reduced by half increases dramatically according to the Mach number, with the cloud in model $c1wind1c$ having moved by $47\, r_{c}$ at $t_{mix}$ (compared to $8\, r_{c}$ for the cloud in model $c1wind1$). This indicates that clouds in higher Mach number winds can travel significant distances before being fully mixed into the flow.

\subsubsection{Cloud shape}
\begin{figure*} 
\centering
\renewcommand{\arraystretch}{3}
  \begin{tabular}{cc}
   \resizebox{70mm}{!}{\includegraphics{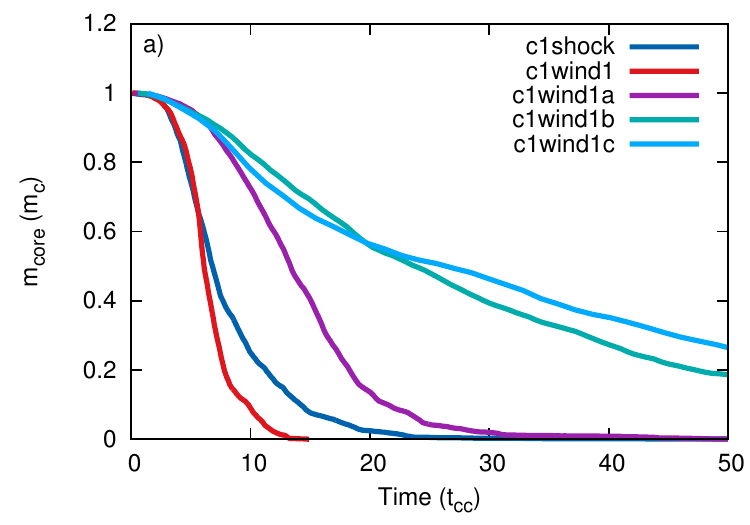}} &
     \resizebox{70mm}{!}{\includegraphics{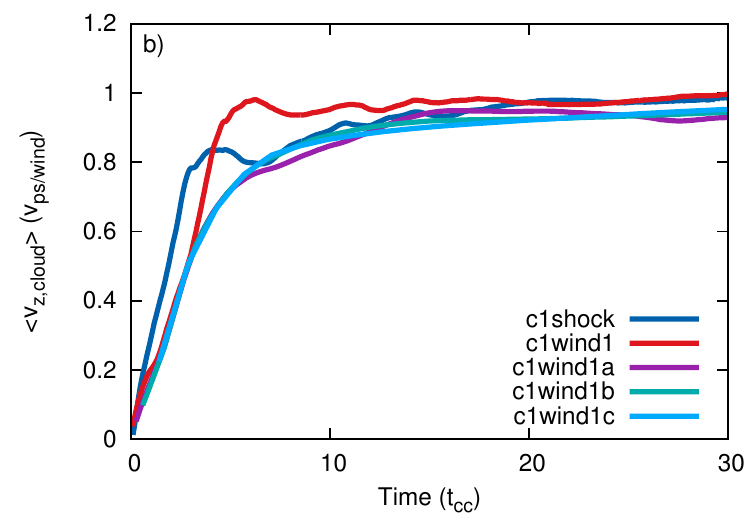}}   
     \multirow{2}{*}{     
    } \\
     \resizebox{70mm}{!}{\includegraphics{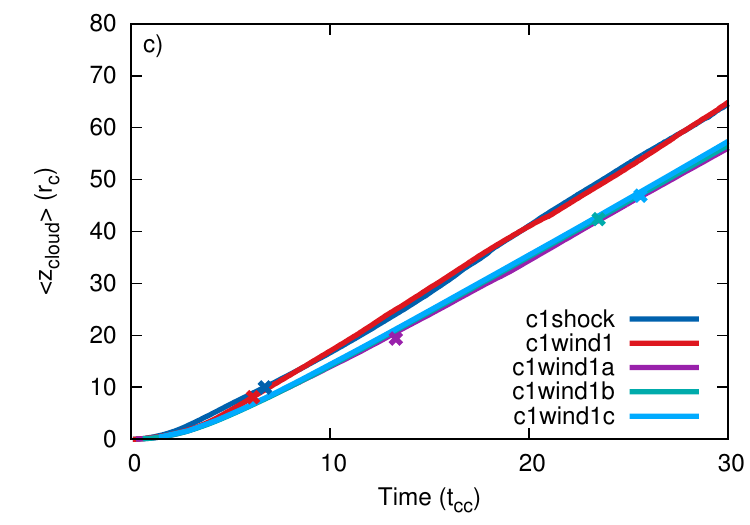}} &
     \resizebox{70mm}{!}{\includegraphics{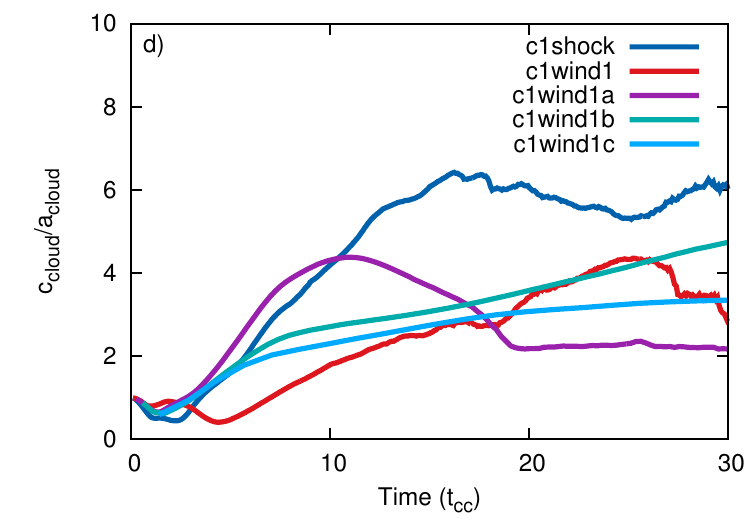}}   \\
      \resizebox{70mm}{!}{\includegraphics{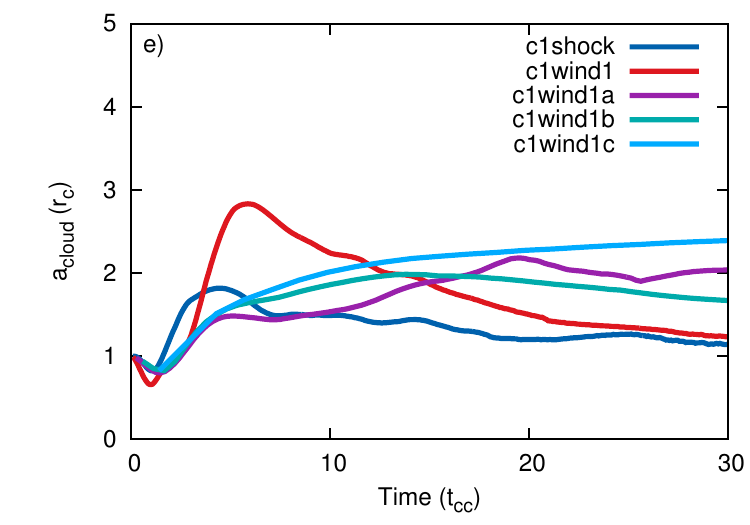}} &
     \resizebox{70mm}{!}{\includegraphics{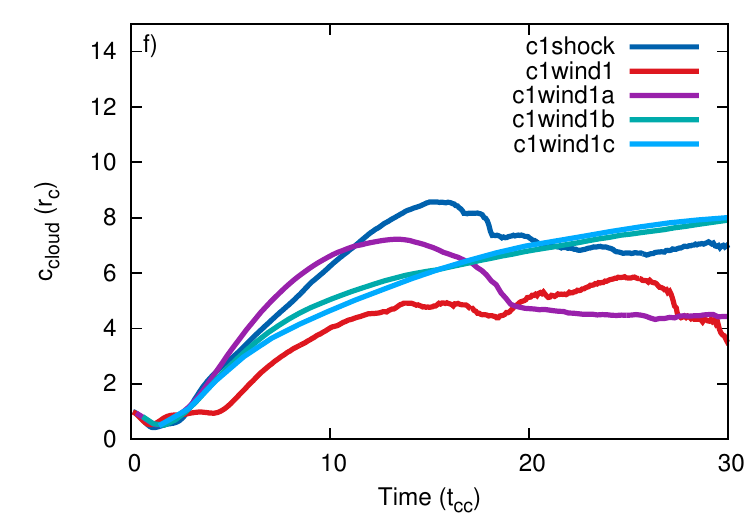}}   \\     
    \end{tabular}
  \caption{Time evolution of (a) the core mass of the cloud, $m_{core}$, (b) the mean velocity of the cloud in the $z$ direction, $\langle v_{z} \rangle$, (c) the centre of mass in the axial direction, $\langle z_{cloud} \rangle$, (d) the ratio of cloud shape in the axial and transverse directions, $c_{cloud}/a_{cloud}$, (e) the effective transverse radius of the cloud, $a_{cloud}$, and (f) the effective axial radius of the cloud $c_{cloud}$. Note that panel (a) shows the evolution on an extended time-scale compared to the other panels. Panel (c) also shows the position of each cloud at $t=t_{mix}$ (indicated by the respective coloured crosses).}
  \label{Fig4}
  \end{figure*}

Figures~\ref{Fig4}(d)-(f) show the time evolution of the effective cloud radii, $a$ and $c$, and their ratio. The radial dimension of the cloud, $a_{cloud}$, decreases slightly during the initial compression phase as the cloud is squeezed in the axial direction, but then increases sharply as the cloud undergoes expansion. Model $c1wind1$ shows the steepest increase, reaching a maximum value for $a_{cloud}$ of $\approx 2.8 \, r_{c}$ at $t=5.9 \,t_{cc}$ as the cloud material is squeezed in the radial direction by the various shocks within and around the cloud, and then decreasing gently as the cloud material is drawn along the axis behind the cloud and gradually mixed into the flow. Model $c1shock$ follows a similar trend, though it reaches its peak expansion of 1.8 $\, r_{c}$ at a slightly earlier time ($t = 4.4 \, t_{cc}$).

The clouds in models $c1wind1b$ and $c1wind1c$ show completely different behaviour, with a more smoothly increasing expansion over time as $M_{wind}$ increases, rather than an initial peak. The cloud in model $c1wind1a$, as noted earlier, displays traits of both behaviours since it shows a slight initial increase before plateauing and then gently increasing again, eventually peaking at an effective radius of $2.2\, r_{c}$ at $t=19.5 \, t_{cc}$.

Since the cloud in simulation $c1shock$ rapidly becomes elongated in the axial direction after the initial compression of the cloud, the values of $c_{cloud}$ and $c_{cloud}/a_{cloud}$ steadily increase over time until $t \approx 17 \, t_{cc}$ when they level out. The cloud in simulation $c1wind1$, in contrast, shows a much less steep increase in $c_{cloud}$ and $c_{cloud}/a_{cloud}$. However, the ratio of cloud shape, $c_{cloud}/a_{cloud}$, shows a much higher value for the cloud in model $c1wind1$, reaching a value of 26 at $t=97\, t_{cc}$ (not shown) while that for model $c1shock$ reaches a high of 8.5 at $t=55\, t_{cc}$. This is in line with \citet{Klein94}, who noted that the combined effect of the lateral expansion associated with the Venturi effect and the axial stretching due to the stripping of material from the side of the cloud led to a much larger cloud aspect ratio for a wind-swept cloud, in comparison to the case of a cloud struck by a shock. 

Similar to the above, models $c1wind1b$ and $c1wind1c$ show a steady increase in both $c_{cloud}$ and $c_{cloud}/a_{cloud}$ (with the plots having very similar profiles for both clouds). In contrast to model $c1wind1$, the clouds in these two simulations have maximum aspect ratios of 11.3 (at $t=221\, t_{cc}$) and 4.4 ($t=214 \, t_{cc}$) (not shown), respectively, which do not follow the behaviour predicted by  \citet{Klein94}. The cloud in model $c1wind1a$ shows different behaviour, again, with an initial peak around $t\approx 10-12\, t_{cc}$ for both $c_{cloud}$ and $c_{cloud}/a_{cloud}$ before levelling off. The peak value for the aspect ratio is 16 at $t=79\, t_{cc}$.

 \subsubsection{Time-scales}  
\begin{figure*} 
\centering
  \begin{tabular}{cc}
   \resizebox{70mm}{!}{\includegraphics{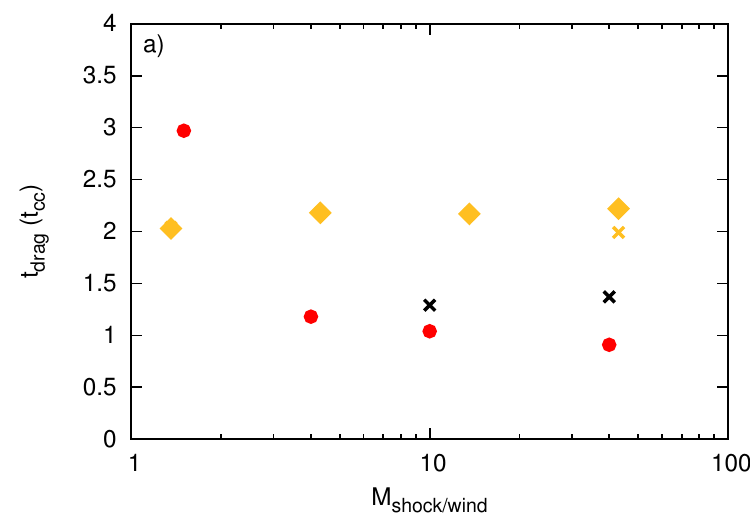}} &
     \resizebox{70mm}{!}{\includegraphics{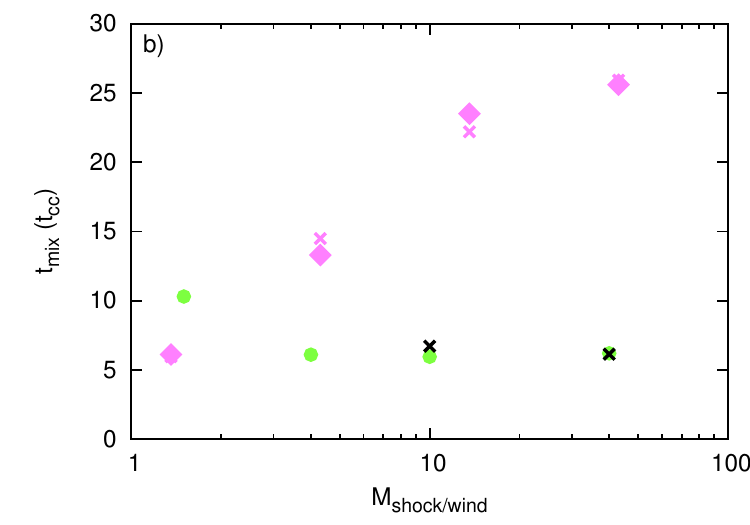}}   \\
    \end{tabular}
  \caption{(a) Cloud drag time, $t_{drag}$, (gold diamonds) and (b) mixing time of the core, $t_{mix}$, (pink diamonds) as a function of the wind Mach number for the wind-cloud simulations. The time-scales for all wind-cloud simulations in this paper which were re-run using the $k$-$\epsilon$ turbulance model are also shown (gold and pink crosses for panels (a) and (b), respectively. Note that these simulations were run at a slightly lower resolution of $R_{64}$). Also shown are the corresponding values as a function of the shock Mach number for shock-cloud simulations with $M_{shock}=10$ and $M_{shock}=40$ (black crosses in each panel), as well as values from the 2D $k$-$\epsilon$ simulations in \citet{Pittard10} for a shock-cloud interaction with $\chi=10$ ($t_{drag}$, red circles; $t_{mix}$, green circles). It should be noted, however, that \citet{Pittard10} used a slightly different definition of the drag time - defined in their paper as the time when the relative cloud velocity had decreased by a factor of $1/e$. This definition provides smaller values of $t_{drag}$ than the calculation used in this paper. In order to compare the two time-scales, we re-calculated our values of $t_{drag}$ for both the shock-cloud simulations where the shock Mach number $M=10$ and $M=40$ and the wind-cloud simulations in accordance with their definition. See Table~\ref{Table3} for values of $t_{drag}$ calculated according to the definition given in \S 2.3 of the current paper. }
  \label{Fig6}
  \end{figure*}

Figure~\ref{Fig6} shows the Mach dependence of $t_{drag}$ and $t_{mix}$. These two time-scales are useful indicators of the evolution and destruction of the cloud. In previous shock-cloud studies \citep[e.g.][]{Pittard10, Pittard16b}, values of $t_{drag}$ and $t_{mix}$ for a given $\chi$ were relatively constant at Mach numbers $>$ 4 (due to Mach scaling), while at lower Mach numbers $t_{drag}$ and $t_{mix}$ both increased sharply. With the wind-cloud simulations, however, we see that the values for $t_{mix}$ increase sharply and nearly linearly (at least for $M_{wind} <10$) as the Mach number increases. The values for $t_{drag}$ for the wind-cloud simulations, meanwhile, are relatively constant within the range $2.0-2.2\, t_{cc}$ \citep[using the definition of $t_{drag}$ found in][]{Pittard10}.\footnote{The calculations performed in \citet{Pittard10} (against which we compare our results in Fig.~\ref{Fig6}) used the $k$-$\epsilon$ turbulence model. In order to ensure that the use of this model had no significant impact on our results, we re-ran our wind simulations using the values for the $k$-$\epsilon$ model employed in \citet{Pittard09, Pittard10} (use of these specific values is important since the strength of turbulent mixing depends on the initial values of $k$ and $\epsilon$ - see \citet{Pittard09} and \citet{Goodson17}). We also calculated a non-$k$-$\epsilon$ model shock-cloud simulation at a shock Mach number of 40. These additional values have been included in Fig.~\ref{Fig6} in order to show clearly the differences between wind-cloud and shock-cloud simulations.} Within this range the cloud in model $c1wind1$ has the lowest value for $t_{drag}$, indicating faster acceleration, and that in model $c1wind1c$ has the highest value (slower acceleration), which fits in with the results of \citet{Scannapieco15} who showed that the acceleration of clouds in galaxy outflows was smaller for higher Mach numbers. While the lack of a shock driven into the back of the cloud in the higher wind Mach number simulations would aid the acceleration of the cloud, it is probable that this effect is superseded by the reduction in the stand-off distance leading to greater compression at the bow shock and a reduction in the normalised wind velocity around the edge of the cloud.  

Figures~\ref{Fig4}(a) and \ref{Fig6}(b) show that the mixing of the core is more efficient at \textit{lower} wind Mach numbers. At lower $M_{wind}$, the growth of KH instabilities is more important and the post-bow shock velocity of the wind around the cloud flanks is greater. At higher $M_{wind}$, $t_{mix}$ levels off at $\simeq 25\, t_{cc}$, indicating that Mach scaling is obtained.

Figure~\ref{Fig6} shows that the ``inviscid'' and ``$k$-$\epsilon$'' models generally have comparable $t_{drag}$ and $t_{mix}$ time-scales, indicating that the level of ``ambient'' turbulence in the latter has little effect on the cloud evolution (higher values are required - see \citet{Pittard09} and \citet{Goodson17}). Instead, one sees much larger differences in $t_{drag}$ and $t_{mix}$ between the shock-cloud and wind-cloud cases, indicating that the \emph{nature} of the background flow is important. 
 
\section{Discussion}
The interaction of both shocks and winds with clouds is of great importance in terms of understanding the nature and evolution of the ISM. Shock-cloud and wind-cloud interactions have been studied numerically but there has been no direct comparison of the two processes, to date. In the following subsections, we discuss two main outcomes of our work, Mach scaling and the long-term survival of the cloud. These have previously been discussed in terms of shock-cloud interactions and we note their importance to wind-cloud studies.

\subsection{Mach scaling}
One of the main results from this study is the presence of Mach scaling. Mach scaling has been discussed in detail in previous shock-cloud studies \citep[see e.g.][]{Klein94, Pittard09, Pittard10}. Briefly, in the strong shock limit, the time evolution of the cloud is independent of the shock Mach number when it is expressed in units of $t/t_{cc} \propto tM$ in the limit $M \rightarrow \infty$. \citet{Klein94} first demonstrated Mach scaling for sharp-edged clouds, with \citet{Nakamura06} producing similar results for clouds with smooth edges. Such studies have been able to demonstrate Mach scaling in the shock-cloud case because the shock Mach numbers used in individual studies have encompassed a large range (e.g. \citet{Klein94} who investigated $M=10-10^{3}$ and \citet{Nakamura06} who used the range $M=1.5-10^{3}$). The same cannot be said for wind-cloud studies. A brief trawl of the literature reveals only a handful of studies where the Mach number of the wind was higher than 10. \citet{Poludnenko04}, in their study of hypersonic radiative bullets, stated that they had used Mach numbers in the range 10-200 but did not go on to discuss the effect of changing the Mach number on the interaction. \cite{Raga07}, who had very similar parameters to those used in the previous study, used a bullet Mach number of 242 which, whilst firmly in the strong shock regime, was not compared to other values of the Mach number. \citet{Pittard05} considered wind Mach numbers of 1 and 20 in their study of multiple clouds embedded in a wind, but did not have a great enough range of values for the Mach number in order to detect Mach scaling. 

Although there are differences in the initial set-up and the physical processes included, our work is perhaps most easily compared to that of \citet{Scannapieco15}, who investigated a range of wind Mach numbers (from 0.5 to 11.4). A key result from these authors was that the mixing time-scale increases with the wind Mach number. However, by extending our investigation to higher wind Mach numbers ($M_{wind}=43.0$ vs. 11.4) we are able to show that the mixing time levels off at high Mach numbers. We believe, therefore, that our paper is the first to demonstrate Mach scaling in a wind-cloud study.

\subsection{Longer survivability of clouds}
In their study, \citet{Scannapieco15} note that clouds embedded in a wind are unable to travel distances of more than $30-40 \, r_{c}$ before being disrupted. We find that clouds can travel $40-50\, r_{c}$ by $t=t_{mix}$, which suggests similarities between our works.

\citet{Scannapieco15} noted that in the hydrodynamic, adiabatic situation only the initial cloud radius determines the distance over which clouds can travel. The authors found that the distances over which the clouds were able to travel would enable them to arrive at a few kpc from the driving region (observations have shown these to be typical distances when clouds are seen in absorption against the starbursting host galaxy \citep[see e.g.][]{Heckman00, Pettini01, Soto12}. These clouds would therefore require a distinct density (as opposed to the cloud mass being smoothed out and mixed into the flow) in order to be observed in this way. Absorption line observations using background galaxies and quasars have in fact revealed that clouds may travel distances on the order of $\approx 100$ kpc or more \citep{Bergeron86, Lanzetta92, Steidel94, Steidel02, Steidel10, Zibetti07, Kacprzak08, Chen10, Tumlinson13, Werk13, Werk14, Peeples14, Turner14}. This is extremely challenging for current theoretical models.

In our study, we find that the cloud in simulation $c1wind1c$, i.e. the simulation with the highest wind velocity and a cloud density contrast of 10, still has significant structure and density at late times (e.g. $100 \,t_{cc}$, when it still has $\approx10\%$ of its core mass; see Fig.~\ref{Fig3}) and that it is able to reach distances of $\simeq 200\, r_{c}$ at this time (see Fig.~\ref{Fig3}). Thus, although our results are still not easily reconciled with observations indicating clouds existing at the 100 kpc distances noted above, they nonetheless show that clouds can survive as distinct structures over much longer distances compared to those presented in \citet{Scannapieco15}. The longer survivability of clouds entrained in a wind may be further enhanced when combined with other effects such as magnetic fields or cooling. 

Figure~\ref{Fig3} shows that the cloud in simulation $c1wind1c$ is not completely destroyed at late times, though its density has dropped below that of the surrounding wind by $t \approx 100\, t_{cc}$ (the bottom panels of Fig.~\ref{Fig3}). Since the bow shock around the cloud is denser than the cloud at this time, preferential detection of the cloud may require that the cloud material has enhanced metallicity relative to the wind \citep[cf.][]{Turner14}.

\section{Summary and Conclusions}
In this paper, we compared the interaction between a shock and a spherical cloud with that of a wind-cloud interaction with similar initial parameters. Our motivation was the lack of any paper in the literature that directly compared these two processes and the general supposition that shock-cloud and wind-cloud interactions were broadly comparable. However, we found there to be subtle, but also significant, differences between the two types of interaction. 

We first compared our wind-cloud simulations against a shock-cloud simulation with $M=10$ and $\chi=10$ ($c1shock$). Our standard wind-cloud simulation ($c1wind1$) has the same cloud completely embedded in a (slightly supersonic) wind with exactly the same properties as the post-shock flow in model $c1shock$. We find that the subsequent behaviour of the external medium differs between the two cases. In the particular case of a marginally supersonic wind, an area of low pressure immediately forms downstream behind the cloud (a feature not present in the shock-cloud case). There are also differences in the morphology of the cloud itself. A cloud engulfed by a marginally supersonic wind undergoes less compression than that struck by a shock, because the flow around the cloud is diffracted in a different way to the shock-cloud case. Finally, there are noticeable differences in the initial transmitted shock between the shock-cloud and wind-cloud simulations; the shock in the former is far flatter in shape whereas that in model $c1wind1$ curves around the edge of the cloud.

As the effective Mach number of the wind increases, the morphological differences between the wind simulations and the shock simulation become more prominent. The cavitation behind the cloud becomes more highly supersonic and elongated. The higher Mach number causes a greater density and pressure jump behind the bow shock, leading to reduced normalised post-bow shock gas velocities around the cloud flank. Because of this, KH instabilities become slightly weaker as $M_{wind}$ increases. Another difference is that clouds in simulations with a high wind Mach number do not experience the formation of transmitted shocks on the axis behind the cloud. In addition to the morphological changes, we also showed that the mixing time increases for increasing $M_{wind}$, which is in contrast to the findings of \citet{Pittard16b} with respect to a shock-cloud interaction. Our simulations also display Mach scaling in the high Mach number limit. The density jump at the bow shock asymptotes to 4.0 (for $\gamma = 5/3$), and the stand-off distance between the bow shock and the centre of the cloud asymptotes to 1.28 $r_{c}$ (again for $\gamma = 5/3$). The morphology of the cloud and the normalised acceleration and mixing time-scales plateau at high Mach numbers. Moreover, we found that clouds embedded in winds with high $M_{wind}$ survived for longer, and travelled over larger distances, compared to the results of the wind-cloud study by \citet{Scannapieco15}.

The models used in this work have several limitations. Firstly, this was a 2D study with imposed axisymmetry. Secondly, we considered only spherical clouds with sharp edges (i.e. our clouds had no distinct core and surrounding envelope but were uniformly dense) and neglected physical processes such as radiative cooling and magnetic fields. Therefore, future comparisons should consider more realistic cloud models and scenarios reflecting a more complex, inhomogenous ISM/intergalactic medium. However, since our work is scale-free our results can be applied to a broad range of problems related to the gas dynamics of the ISM. A follow-up paper to the present study will compare shock-cloud and wind-cloud interactions where the cloud density contrast is higher.

\section*{Acknowledgements}
We would like to thank the referee, Alejandro Raga, for constructive comments that helped to clarify and generally improve the manuscript. This work was supported by the Science \& Technology Facilities Council [Research Grants ST/L000628/1 and ST/M503599/1]. We thank S. Falle for the use of the \textsc{mg} hydrodynamics code used to calculate the simulations in this work. The calculations used in this paper were performed on the DiRAC Facility which is jointly funded by STFC, the Large Facilities Capital Fund of BIS, and the University of Leeds. The data associated with this paper are openly available from the University of Leeds data repository. \url{http://doi.org/10.5518/120}


\begin{thebibliography}{}

\bibitem[\protect\citeauthoryear{Al\={u}zas et al.}{2012}] {Aluzas12} Al\={u}zas R., Pittard J. M., Hartquist T. W., Falle S. A. E. G., Langton R., 2012, MNRAS, 425, 2212
\bibitem[\protect\citeauthoryear{Al\={u}zas et al.}{2014}] {Aluzas14} Al\={u}zas R., Pittard J. M., Falle S. A. E. G., Hartquist T. W., 2014, MNRAS, 444, 971
\bibitem[\protect\citeauthoryear{Banda-Barrag\'{a}n et al.} {2016}] {Banda16} Banda-Barrag\'{a}n W. E., Parkin E. R., Crocker R. M., Federrath C., Bicknell G. V., 2016, MNRAS, 455, 1309
\bibitem[\protect\citeauthoryear{Bergeron} {1986}] {Bergeron86} Bergeron J., 1986, A\&A, 155, L8 
\bibitem[\protect\citeauthoryear{Bruhweiler et al.} {2010}] {Bruhweiler10} Bruhweiler F. C., Ferrero R. F., Bourdin M. O., Gull T. R., 2010, ApJ, 719, 1872
\bibitem[\protect\citeauthoryear{Chen et al.} {2010}] {Chen10} Chen H. -W., et al., 2010, ApJ, 714, 1521
\bibitem[\protect\citeauthoryear{Cooper et al.} {2008}] {Cooper08} Cooper J. L., Bicknell G. V., Sutherland R. S., Bland-Hawthorn J., 2008, ApJ, 674, 157
\bibitem[\protect\citeauthoryear{Cooper et al.} {2009}] {Cooper09} Cooper J. L., Bicknell G. V., Sutherland R. S., Bland-Hawthorn J., 2009, ApJ, 703, 330
\bibitem[\protect\citeauthoryear{Cowie et al.} {1981}] {Cowie81} Cowie L. L., McKee C. F., Ostriker J. P., 1981, ApJ, 247, 908
\bibitem[\protect\citeauthoryear{Dyson et al.} {2002}] {Dyson02} Dyson J. E., Arthur S. J., Hartquist T. W., 2002, A\&A, 390, 1063
\bibitem[\protect\citeauthoryear{Elmegreen \& Scalo}{2004}] {Elmegreen04} Elmegreen B. G., Scalo J., 2004, ARA\&A, 42, 211
\bibitem[\protect\citeauthoryear{Falle}{1991}]{Falle91} Falle S. A. E. G., 1991, MNRAS, 250, 581
\bibitem[\protect\citeauthoryear{Farris \& Russell} {1994}] {Farris94} Farris M. H., Russell C. T., 1994, JGR, 99, 17681
\bibitem[\protect\citeauthoryear{Fragile et al.} {2004}] {Fragile04} Fragile P. C., Murray S. D., Anninos P., van Breugel W., 2004, ApJ, 604, 74
\bibitem[\protect\citeauthoryear{Goldsmith \& Pittard} {2016}] {Goldsmith16} Goldsmith K. J. A., Pittard J. M., 2016, MNRAS, 461, 578
\bibitem[\protect\citeauthoryear{Goodson et al.} {2017}] {Goodson17} Goodson M.D., Heitsch D., Eklund K., Williams V. A., 2017, MNRAS, 468, 318
\bibitem[\protect\citeauthoryear{Gregori et al.} {1999}] {Gregori99} Gregori G., Miniati F., Ryu D., Jones T. W., 1999, ApJ, 527, L113
\bibitem[\protect\citeauthoryear{Gregori et al.} {2000}] {Gregori00} Gregori G., Miniati F., Ryu D., Jones T. W., 2000, ApJ, 543, 775
\bibitem[\protect\citeauthoryear{Hansen et al.} {2007}] {Hansen07} Hansen J. F., Robey H. F., Klein R. I., Miles A. R., 2007, ApJ, 662, 379
\bibitem[\protect\citeauthoryear{Heckman et al.} {2000}] {Heckman00} Heckman T., Lehnert M. D., Strickland D. K., Lee A., 2000, ApJS, 129, 493
\bibitem[\protect\citeauthoryear{Hennebelle \& Falgarone} {2012}] {Hennebelle12} Hennebelle P., Falgarone E., 2012, A\&AR, 20, 55
\bibitem[\protect\citeauthoryear{Johansson \& Ziegler} {2013}] {Johansson13} Johansson E. P. G., Ziegler U., 2013, ApJ, 766, 45
\bibitem[\protect\citeauthoryear{Jones et al.} {1996}] {Jones96} Jones T. W., Ryu D., Tregillis I. L., 1996, ApJ, 473, 365
\bibitem[\protect\citeauthoryear{Kacprzak et al.} {2008}] {Kacprzak08} Kacprzak G. G., et al., 2008, AJ, 135, 922
\bibitem[\protect\citeauthoryear{Klein et al.} {1994}] {Klein94} Klein R. I., McKee C. F., Colella P., 1994, ApJ, 420, 213
\bibitem[\protect\citeauthoryear{Klein et al.} {2003}] {Klein03} Klein R. I., Bundil K. S., Perry T. S., Bach D. R., 2003, ApJ, 583, 245
\bibitem[\protect\citeauthoryear{Koo et al.} {2001}] {Koo01} Koo B-C., Rho J., Reach W. T., Jung J., Mangum J. G., 2001, ApJ, 552, 175
\bibitem[\protect\citeauthoryear{Lanzetta \& Bowen} {1992}] {Lanzetta92} Lanzetta K. M., Bowen D. V., 1992, ApJ, 391, 48
\bibitem[\protect\citeauthoryear{Li et al.}{2013}] {Li13} Li S., Frank A., Blackman E. G., 2013, ApJ, 774, 133
\bibitem[\protect\citeauthoryear{Mac Low et al.} {1994}] {MacLow94} Mac Low M.-M., McKee C. F., Klein R. I., Stone J. M., Norman M. L., 1994, ApJ, 433, 757
\bibitem[\protect\citeauthoryear{Mac Low \& Klessen} {2004}] {MacLow04} Mac Low M.-M., Klessen R., 2004, Rev. Mod. Phys., 76, 125
\bibitem[\protect\citeauthoryear{Marcolini et al.} {2005}] {Marcolini05} Marcolini A., Strickland D. K., D'Ercole A., Heckman T. M., Hoopes C. G., 2005, MNRAS, 362, 626
\bibitem[\protect\citeauthoryear{Martin} {2005}] {Martin05} Martin C. L., 2005, ApJ, 621, 227
\bibitem[\protect\citeauthoryear{Martin et al.} {2012}] {Martin12} Martin C. L., et al., 2012, ApJ, 760, 127
\bibitem[\protect\citeauthoryear{McCourt et al.} {2015}] {McCourt15} McCourt M., O'Leary R. M., Madigan A. -M., Quataert E., 2015, MNRAS, 449, 2
\bibitem[\protect\citeauthoryear{McKee \& Ostriker} {1977}] {McKee77} McKee C. F., Ostriker J. P., 1977, ApJ, 218, 148
\bibitem[\protect\citeauthoryear{McKee \& Ostriker} {2007}] {McKee07} McKee C. F., Ostriker E. C., 2007, ARA\&A, 45, 565
\bibitem[\protect\citeauthoryear{Mellema et al.} {2002}] {Mellema02} Mellema G., Kurk J. D., R\"{o}ttgering H. J. A., 2002, A\&A, 395, L13
\bibitem[\protect\citeauthoryear{Miniati et al.} {1999}] {Miniati99} Miniati F., Jones T. W., Ryu D., 1999, ApJ, 517, 242
\bibitem[\protect\citeauthoryear{Nakamura et al.} {2006}] {Nakamura06} Nakamura F., McKee C. F., Klein R. I., Fisher R. T., 2006, ApJ, 164, 477
\bibitem[\protect\citeauthoryear{Orlando et al.} {2005}] {Orlando05} Orlando S., Peres G., Reale F., Bocchino F., Rosner R., Plewa T., Siegel A., 2005, A\&A, 444, 505
\bibitem[\protect\citeauthoryear{Orlando et al.} {2008}] {Orlando08} Orlando S., Bocchino F., Reale F., Peres G., Pagano P., 2008, ApJ, 678, 274
\bibitem[\protect\citeauthoryear{Padoan et al.} {2014}] {Padoan14} Padoan P., et al., 2014, in Protostars and Planets VI, Henrik Beuther, Ralf
S. Klessen, Cornelis P. Dullemond, and Thomas Henning (eds.), University of Arizona Press, Tucson, pp 77
\bibitem[\protect\citeauthoryear{Peeples et al.} {2014}] {Peeples14} Peeples M. S., et al., 2014, ApJ, 786, 54
\bibitem[\protect\citeauthoryear{Peretto et al.} {2012}] {Peretto12} Peretto N., et al., 2012, A\&A, 541, A63
\bibitem[\protect\citeauthoryear{Pettini et al.} {2001}] {Pettini01} Pettini M., et al., 2001, ApJ, 554, 981
\bibitem[\protect\citeauthoryear{Pittard \& Goldsmith}{2016}] {Pittard16a} Pittard J. M., Goldsmith K. J. A., 2016, MNRAS, 458, 1139
\bibitem[\protect\citeauthoryear{Pittard \& Parkin}{2016}] {Pittard16b} Pittard J. M., Parkin E. R., 2016, MNRAS, 457, 4470
\bibitem[\protect\citeauthoryear{Pittard et al.} {2003}] {Pittard03} Pittard J. M., Arthur S. J., Dyson J. E., Falle S. A. E. G., Hartquist T. W., Knight M. I., Pexton M., 2003, A\&A, 401, 1027
\bibitem[\protect\citeauthoryear{Pittard et al.} {2005}] {Pittard05} Pittard J. M., Dyson J. E., Falle S. A. E. G., Hartquist T. W., 2005, MNRAS, 361, 1077
\bibitem[\protect\citeauthoryear{Pittard et al.} {2009}] {Pittard09} Pittard J. M., Falle S. A. E. G., Hartquist T. W., Dyson J. E., 2009, MNRAS, 394, 1351
\bibitem[\protect\citeauthoryear{Pittard et al.} {2010}] {Pittard10} Pittard J. M., Hartquist T. W., Falle S. A. E. G., 2010, MNRAS, 405, 821
\bibitem[\protect\citeauthoryear{Poludnenko et al.}{2002}]{Poludnenko02} Poludnenko A. Y., Frank A., Blackman E. G., 2002, ApJ, 576, 832
\bibitem[\protect\citeauthoryear{Poludnenko et al.}{2004}]{Poludnenko04} Poludnenko A. Y., Frank A., Mitran S., 2004, ApJ, 613, 387
\bibitem[\protect\citeauthoryear{Raga et al.} {2005}] {Raga05} Raga A., Steffen W., Gonz\'{a}lez R., 2005, Revista Mexicana, 41, 45
\bibitem[\protect\citeauthoryear{Raga et al.} {2007}] {Raga07} Raga A. C., Esquivel A., Riera A., Vel\'{a}zquez P. F., 2007, ApJ, 668, 310
\bibitem[\protect\citeauthoryear{Rogers \& Pittard} {2013}] {Rogers13} Rogers H., Pittard J. M., 2013, MNRAS, 431, 1337
\bibitem[\protect\citeauthoryear{Rupke et al.} {2002}] {Rupke02} Rupke D. S., Veilleux S., Sanders D. B., 2002, ApJ, 570, 588
\bibitem[\protect\citeauthoryear{Sales et al.}{2010}] {Sales10} Sales L. V., Navarro J. F., Schaye J., Dalla Vecchia C., Springel V., Booth C. M., 2010, MNRAS, 409, 1541
\bibitem[\protect\citeauthoryear{Scalo \& Elmegreen}{2004}] {Scalo04} Scalo J., Elmegreen B. G., 2004, ARA\&A, 42, 275
\bibitem[\protect\citeauthoryear{Scannapieco \& Br\"{u}ggen}{2015}] {Scannapieco15} Scannapieco E., Br\"{u}ggen M., 2015, ApJ, 805, 158
\bibitem[\protect\citeauthoryear{Schiano et al.}{1995}] {Schiano95} Schiano V. R., Christiansen W. A., Knerr J. M., 1995, ApJ, 439, 237
\bibitem[\protect\citeauthoryear{Schneider et al.} {2006}] {Schneider06} Schneider N., et al., 2006, A\&A, 458, 855
\bibitem[\protect\citeauthoryear{Shin et al.} {2008}] {Shin08} Shin M.-S., Stone J. M., Snyder G. F., 2008, ApJ, 680, 336
\bibitem[\protect\citeauthoryear{Soto \& Martin} {2012}] {Soto12} Soto K. T., Martin C. L., 2012, ApJ, 203, 3
\bibitem[\protect\citeauthoryear{Steidel et al.} {1994}] {Steidel94} Steidel C. C., Dickinson M., Persson S. E., 1994, ApJL, 437, L75
\bibitem[\protect\citeauthoryear{Steidel et al.} {2002}] {Steidel02} Steidel C. C., et al., 2002, ApJ, 570, 526
\bibitem[\protect\citeauthoryear{Steidel et al.} {2010}] {Steidel10} Steidel C. C., et al., 2010, ApJ, 717, 289
\bibitem[\protect\citeauthoryear{Stone \& Norman} {1992}] {Stone92} Stone J. M., Norman M. L., 1992, ApJ, 390, L17
\bibitem[\protect\citeauthoryear{Tumlinson et al.} {2013}] {Tumlinson13} Tumlinson J., et al., 2013, ApJ, 777, 59
\bibitem[\protect\citeauthoryear{Turner et al.} {2014}] {Turner14} Turner M. L., Schaye J., Steidel C. C., Rudie G. C., Strom A. L., 2014, MNRAS, 445, 794
\bibitem[\protect\citeauthoryear{Wareing et al.} {2017}] {Wareing17} Wareing C., Pittard J. M, Falle S. A. E. G., 2017, MNRAS. 465, 2757
\bibitem[\protect\citeauthoryear{Werk et al.} {2013}] {Werk13} Werk J. K., et al., 2013, ApJS, 204, 17
\bibitem[\protect\citeauthoryear{Werk et al.} {2014}] {Werk14} Werk J. K., et al., 2014, ApJ, 792, 8
\bibitem[\protect\citeauthoryear{Westmoquette et al.} {2010}] {Westmoquette10} Westmoquette M. S., Slavin J. D., Smith L. J., Gallagher J. S., 2010, MNRAS, 402, 152
\bibitem[\protect\citeauthoryear{White \& Long} {1991}] {White91} White R. L., Long K. S., 1991, ApJ, 373, 543
\bibitem[\protect\citeauthoryear{Xu \& Stone} {1995}] {Xu95} Xu J., Stone J. M., 1995, ApJ, 454, 172
\bibitem[\protect\citeauthoryear{Yirak et al.} {2010}] {Yirak10} Yirak K., Frank A., Cunningham A. J., 2010, ApJ, 722, 412
\bibitem[\protect\citeauthoryear{Zibetti et al.} {2007}] {Zibetti07} Zibetti S., et al., 2007, ApJ, 658, 161
\end{thebibliography}
\end{document}